\documentstyle[12pt,svoss2,psfig]{book}

\begin{document}

\title{Causation, existence, and creation in space-times with non-trivial topology}

\author{Gustavo E. Romero}

\affil{Instituto Argentino de Radioastronom\'{\i}a, C.C.5, (1894)
Villa Elisa, Buenos Aires, Argentina. {\sl Email:
romero@irma.iar.unlp.edu.ar}}

\begin{abstract}
The Kalam Cosmological Argument is perhaps the most solid and
widly discussed argument for a caused creation of the universe.
The usual objections to the argument mainly focus on the second
premise. In this paper we discuss the dependency of the first
premise on the topological structure of the space-time manifold
adopted for the underlying cosmological model. It is shown that in
chronology-violating space-times the first premise is also
violated. The chronology-violation, in turn, requires a massive
violation of the so-called energy conditions which could have
observational effects that are briefly discussed here. Hence,
astronomical observations could be relevant for the validity of
the metaphysical argument. In this sense, it is possible to talk
of ``observational theology".
\end{abstract}

\keywords{space-time, existence, creation, cosmology}


\section{Introduction}

The so-called Kalam Cosmological Argument (e.g. Craig 1979) is a
version of the classical cosmological argument based on some
medieval Islamic arguments against the infinitude of the past. In
modern syllogistic form it can be formulated as follows:

\begin{enumerate}
\item Whatever begins to exist has a cause of its existence. \item
The universe began to exist. \item Therefore, the universe has a
cause of its existence.
\end{enumerate}

It has been argued that the first premise is a fundamental
metaphysical principle which cannot be intelligibly denied and
that the second premise is supported by modern cosmology, in such
a way that the conclusion of the argument is true (Craig 1979,
Craig \& Smith 1993). These contentions have been discussed in
recent years by several philosophers, notably Adolf G\"unbaum, who
argued that the Big Bang model does not support the second premise
(e.g. Gr\"unbaum 1989, 1990, 1991, 2000, and some replies in Craig
1991 and 1992). The first premise, on the contrary, has not been
considered controversial except from the point of view of quantum
mechanics (see the discussions in Craig \& Smith 1993).

In this paper we shall argue that the validity of the first
premise depends on the topology of space-time manifold adopted for
the cosmological model. Multiple connected space-times can be
compatible with objects that obey all physical laws but violate
the first premise of the Kalam Cosmological Argument. Some
semantic comments are in order first to clarify the meaning of the
expression ``to begin to exist''.

\section{A semantical note}

Craig (1992) attributes to Gr\"unbaum the implicit use of the
following definition:\\

{ ``$x$ begins to exist''$=$def. ``$x$ exists at time $t$ and
there are instants of time immediately prior to $t$ at which $x$
does not exist''. \label{def1}}\\

This definition is objected because it is difficult to accept that
the existence of $x$ at $t$ can entail the existence of temporal
instants prior to $t$. Admittedly, in the context of a relational
theory of space-time (e.g. Perez-Bergliaffa, Romero \& Vucetich
1998) the requirement of the existence of moments prior to $t$ is
nothing else than the requirement of the existence of objects
other than $x$ {\sl before} $x$. Such a definition, then, is not
adequate to the discussion of the origin of the system formed by
{\sl all} things, i.e. the universe. Craig, in turn, proposes:\\

{\rm ``$x$ begins to exist''$=$def. ``$x$ exists at time $t$ and
there are no instants of time immediately prior to $t$ at which
$x$ exists''. \label{def2}}\\

This allows for a beginning of time itself and is apparently apt
for a discussion on the beginning of the universe. But it has the
problem of demanding a sharp edge for the existence of $x$.
Anything created by an evolutionary process lasting a finite time
interval is excluded. Let us consider, for example, the Mankind.
It certainly exists now and it certainly did not exist 50 million
years ago, but can we point out an instant $t$ at which it did
existed and an immediately prior instant at which it did not?. Not
only biological counterexamples are possible, but we can think
also in most physical systems, like a star or a molecular cloud,
which are formed by a slow transition from a previous state.

In order to remove this problem we propose:\\

{ ``$x$ begins to exist''$=$def. ``$x$ exists at time $t$ and
there is a time interval $\Delta t \geq 0$ such that there are no
instants of time immediately prior to $t-\Delta t$ at which $x$
exists''.  \label{def3}}\\

For $\Delta t = 0$ we recover Craig's definition. In what follows
we shall understand ``to begin to exist'' in the sense of this
latter definition.

\section{Chronology-violating space-times and self-existent objects}

A relativistic space-time is represented by a four-dimensional
manifold $M$ equipped with a Lorentzian metric $g_{ab}$. The
General Theory of Relativity requires the manifold to be
continuous and differentiable but not specific constraints are
imposed on the details of its topology. Usually, simply connected
manifolds are considered, but multiply connected ones cannot be
ruled out only on a priori grounds.

In recent years there has been a sustained interest in multiple
connected space-times, also called wormhole space-times,
originated in the fact that close timelike curves (CTCs) naturally
appear in them (e.g. Morris, Thorne \& Yurtsever 1988, Thorne
1992). These curves represent the world lines of any physical
system in a temporally orientable space-time that, moving always
in the future direction, ends arriving back at some point of its
own past. Any space-time with CTCs is called a
chronology-violating space-time. Objections to the formation of
CTCs in the real universe had been formulated by a number of
scientists, most notably by Hawking (1992), but in the absence of
a theory of quantum gravity the possibility of wormholes in
space-time cannot be ruled out (see the discussions and references
in Earman 1995a, Romero \& Torres 2001, and Nahin 1999).

One of the most strange implications of chronology-violating
space-times is the possibility of an ontology with self-existent
objects. These are physical systems ``trapped'' in CTCs. Romero \&
Torres (2001), who have discussed these systems in depth, give the
following toy-example to illustrate the nature of such objects:\\

{\it Suppose that, in a space-time where CTCs exist, a time
traveler takes a ride on a time machine carrying a book with her.
She goes back to the past, forgets the book in -what will be- her
laboratory, and returns to the future. The book remains then
hidden until the time traveler finds it just before starting her
time trip, carrying the book with her. }\\

The book in question is a self-existent object: it exists at a
given $t$, there exists $\Delta t \geq 0$ such that the object
does not exist at $t-\Delta t$, but, however, there is not an
external cause of its existence. The self-existent object is {\sl
just a feature of space-time itself}, it is not either created or
destroyed {\sl in} space-time. Such objects clearly violate the
first premise of the Kalam Cosmological Argument.

It is very important to emphasize that, despite that the
self-existent objects have not a cause of their existence, they do
not violate causality. In fact, since their space-time history is
a continuous closed curve, {\sl their physical state at every time
$t$ is casually linked to a previous state}. In this way, these
objects are not causally created, but they have a finite existence
in the sense that they exist during a finite time interval, and
their existence does not violate strict causality.

Romero \& Torres (2001) have argued against an ontology of
self-existent objects invocating a full Principle of
Self-Consistency for {\sl all} laws of nature. This principle,
which is used to dissolve the so-called ``paradoxes'' of time
travel (Earman 1995b, Nahin 1999), can be stated as:\\

{\it The laws of nature are such that any local solution of their
equations that represents a feature of the real universe must be
extendible to a global solution}.\\

Romero and Torres suggest that this principle is a {\sl
metanomological} statement (see Bunge 1961) that enforces the
harmony between local and global affairs in space-time. By
including thermodynamics in the consistency analysis of the motion
of macroscopic systems through wormhole space-times, they have
shown that {\sl non-interacting} self-existent objects are not
possible in the real universe because energy degradation along the
CTC results in non-consistent histories.

Notwithstanding these objections, the development of consistent
histories remains an open possibility for isolated systems where
entropy cannot be defined (e.g. single particles) and for
interacting systems where their energy degradation is exactly
compensated by external work made upon them (Lossev \& Novikov
1992). Hence, if CTCs actually occur in the universe, there seems
to be no form to avoid the possibility of at least some types of
self-existent objects.

Very recently, J. Richard Gott III and Li-Xin Li (1998) have even
proposed that the universe itself could be a self-existent object.
Form a philosophical point of view, this would be a violation of
both premises of the Kalam Cosmological Argument with a single
counterexample. As far as it can be seen, the work by Gott and Li
is consistent with the Big Bang paradigm. They only require the
existence of a multiply connected space-time with a CTC region
beyond the original inflationary state.

A key point for the validity of the first premise of the Kalam
Cosmological Argument is that the space-time in the real universe
must be described by a simply connected manifold, with no CTCs
\footnote{Formally, CTCs are possible even in simply connected
space-times, but these kind of solutions of Einstein field
equations, like the classical G\"odel (1949) rotating universe,
are thought to be not applicable to the real world.}. Otherwise,
the presence of objects that have ``began to exist'' without
external cause but notwithstanding are subject to causality cannot
be excluded. We have, then, two possibilities in order to explore
the validity of the first premise in the context of its dependency
on the underlying topology of space-time: 1) we can try to prove,
from basic physical laws, that CTCs cannot be formed in the real
universe (i.e. we can try to find out a mechanism to enforce
chronology protection), or 2) we can inquire about the
observational signatures of wormhole structures in space-time and
try to test through observations the hypothesis that natural
wormholes actually do exist. The first option requires a full
theory of Quantum Gravity, something that is beyond our present
knowledge. The second approach is being already explored by some
scientists.

\section{Observational signatures of WEC-violating matter}

Macroscopic and static wormhole structures as those necessary to
allow the formation of CTCs require that the average null energy
condition (ANEC) be violated in the wormhole throat. This
condition is part of the so-called energy conditions of Einstein
gravity, which are very general hypothesis designed to provide as
much information as possible on a wide variety of physical systems
{\sl without} specifying a particular equation of state. These
conditions {\sl are not} proved from basic principles, they are
just conjectures, which can be very useful in some contexts.
However, many violating systems are known, including the universe
itself (see Visser 1996).

The energy conditions violated by traversable wormhole can be put
in terms of the stress-energy tensor of the matter
threading the wormhole as $ \rho+p\geq 0 $, where $\rho$ is the
energy density and $p$ is the total pressure. This implies also a
violation of the so-called weak energy condition --WEC--
($\rho\geq 0\;\wedge\;\rho+p\geq 0$; see Visser 1996 for details,
also Morris and Thorne 1988). Plainly stated, all this means that
the matter threading the wormhole must exert gravitational
repulsion in order to stay stable against collapse. If natural
wormholes exist in the universe (e.g. if the original topology
after the Big-Bang was multiply connected), then there should be
observable signatures of the interactions between matter with
negative energy density with the normal matter.

At astronomical level the most important observational consequence
of the existence of natural wormholes is gravitational lensing of
background sources (Cramer et al. 1995, Torres et al. 1998; Eiroa
et al. 2001, Safonova et al. 2001). There are very specific
features produced by chromaticity effects in lensing of extended
sources that could be used to differentiate events produced by
wormholes from those of other objects (Eiroa et al. 2001). In the
wormhole microlensing case there are two intensity peaks during
each event separated by an umbra region. On the contrary, in the
normal case there is a single, time-symmetric peak. In addition,
in the wormhole case it can be shown that there is a spectral
break that is not observed in the usual case (Eiroa et al. 2001
for details).

Also, the macrolensing effects upon a background field of galaxies
produced by large-scale violations of the energy conditions are
observationally distinguishable from the normal macrolensing by
either dark or luminous matter concentrations (see Safonova et al.
2001 for complete numerical simulations of macrolensed galaxy
fields). In particular, it can be shown that for positive mass we
see concentric arcs, whereas for negative energy densities we have
filamentary features projected from the center.

The above examples are enough to illustrate the kind of
observational effects that can be expected in an universe with
multiple connected topology. Whether such space-time wormholes
actually exist in our universe is something that has to be found
yet.

The mere existence of a multiple connected topology for space-time
does not warrant, by itself, the violation of the first premise of
the Kalam Cosmological Argument. But it makes possible the
formation of CTCs and non-cronal situations in that space-time,
hence opening the possibility of an ontology with self-existent
objects. This implies that the universality of the premise can be
objected even at a macroscopic level, without resorting to quantum
considerations.

\section{Conclusions: Theology meets experiment}

The first premise of the Kalam Cosmological Argument, namely that
``whatever that begins to exist has a cause of its existence'', is
not a self-evident, universally valid statement as it is usually
accepted. We have shown that the truth value of the premise is
dependent on some basic characteristics of the space-time manifold
that represents the real universe. In particular, multiple
connected space-times can accommodate objects that exist by
themselves, without external cause, but also without any local
violation of causality. These objects ``begin to exist'' in
accordance to even the most restrictive definitions given in
Section 2.

Since the connectivity of space-time can be probed through
astronomical observations (see Anchordoqui et al. 1999 for an
example of these observational studies), the validity of the Kalam
Cosmological Argument can be tested by the scientific method. Not
only the second premise, which uses to be discussed in the light of
the Big Bang cosmology, but also the first premise of the argument
is susceptible to experimental test. It is in this more extended
sense that in the Kalam Cosmological Argument we can say that
theology meets experiment.

\acknowledgments

I thank Diego F. Torres and Santiago E. Perez-Bergliaffa for many
illuminating discussions on space, time, and causality along many
years. This work has been supported by
the agencies CONICET and ANPCT (PICT 03-04881) as well as by funds
granted by Fundaci\'{o}n Antorchas.

\end{document}